\documentclass[pra,showpacs,twocolumn,a4paper,showkeys,
superscriptaddress]{revtex4}

\usepackage{times}
\usepackage{graphicx}
\usepackage{amsmath, amssymb}

\usepackage[T1]{fontenc}
\usepackage{textcomp}
\usepackage{mathptmx}

\newcommand{\bra}[1]{\left\langle#1\right|}
\newcommand{\ket}[1]{\left|#1\right\rangle}

\begin{document}
\title{A number filter for matter-waves} 
\author{G. Nandi}
\affiliation{Institut f\"ur Quantenphysik, Universit\"at Ulm, D-89069 Ulm, Germany}  
\author{A. Sizmann} 
\affiliation{Ludwig-Maximilians-Universit\"at  
M\"unchen, D-80539 M\"unchen} 
\author{J. Fort\'agh}
\affiliation{Physikalisches Institut der Universit\"at T\"ubingen, D-72076
  T\"ubingen, Germany} 
\author{R. Walser}
\affiliation{Institut f\"ur Quantenphysik, Universit\"at Ulm, D-89069 Ulm, Germany}  
\email{Reinhold.Walser@uni-ulm.de}
\date{\today}

\begin{abstract}
  In current Bose-Einstein condensate experiments, the shot-to-shot variation
  of atom number fluctuates up to 10$\%$.  In here, we present a procedure to
  suppress such fluctuations by using a nonlinear $p-\pi-\bar{p}$ matter wave
  interferometer for a Bose-Einstein condensate with two internal states and a
  high beam-splitter asymmetry ($p,\bar{p}\neq 0.5$). We analyze the situation
  for an inhomogeneous trap within the Gross-Pitaevskii mean-field theory, as
  well as a quantum mechanical Josephson model, which addresses complementary
  aspects of the problem and agrees well otherwise.
\end{abstract}

 \maketitle

\section{Introduction}

Nonlinear optical wave propagation is known to give rise to chaos, spectral
and temporal distortion or an amplification of noise \cite{agrawal}. This
fundamentally limits the signal-to-noise ratio in state-preparation
experiments or measurements.  For example, the nonlinearity constrains the
capacity of optical communication systems \cite{Mitra2001}, or the resolution
of a gravitational-wave interferometer \cite{Jaekel1990}, where the momentum
transfer to the mirror produces an intensity-dependent phase shift.  However,
optical nonlinearities are also capable of wave-packet self-stabilization and
of a phase-sensitive reduction of noise.  Second- and third-order
nonlinearities, and especially the Kerr effect in optical fibers, produce
energy stabilization and a noise reduction below the standard quantum limit
\cite{Ritze1979,kaplan81,Kitagawa1986,
  Schmitt1998,Werner1998b,wood88,Sizmann1999,Levandovsky1999} in various
experimental configurations \cite{Bachor2003}.

The past decade of matter-wave physics has also shown remarkable similarities
with the development of quantum optics in the 60's. A lucent description of
this parallelism of quantum optics \cite{schleich01} and atomic matter waves
is found in \cite{moelmer05}.  Starting from the seminal measurement of
spatial coherence in normal and degenerate gases
\cite{shimitsu96,ketterleinterference,CohenTannoudji97,esslinger02}, the field
has eventually progressed to study density fluctuations in trapped, three
dimensional Bose-Einstein condensates (BEC)
\cite{Orzel01,Raizen2005a,westbrook05} and fermionic gases \cite{westbrook07}.
By reducing dimensionality via geometric confinement in planar traps,
one-dimensional traps, in optical lattices or on atomic chips \cite{fortagh07}
the field has now been opened to a plethora of condensed matter phenomena
\cite{jaksch98,goerlitz01,bloch02,esslinger04,bloch05,ertmer03,
  walser04,Weiss05a}. While the perfection of communication quality is the key
issue for optics today, the main application for cold atomic matter waves is
quantum metrology and sensing. Reaching the quantum limit and surpassing it
with matter-waves is a major research direction
\cite{ramsey90,ueda91,holland93,wineland94,berman,becmugrav06,nandi06,
  kajari04,ertmer06,kasevich07a,kasevich07b}.  In this context, atoms or ions
prove to be more flexible than light, as we can control many-particle
entanglement and exploit different quantum statistics
\cite{moelmer99,polzik01,zoller_nature01}.

The statistical ensembles that are generated in most of the current
experiments are never of the quality as theoretically envisaged. In
particular, most of the current BEC experiments face a shot-to-shot variation
of particle number $N$ of about 10$\%$. This is primarily due to technical
uncertainties in the evaporation procedure.  If each individual BEC
realization would be characterized by a pure Fock state $\ket{\Psi_N}$, then
the uncertainty ${\cal P}_N$ in atom number will lead to an mixed state
ensemble with a density operator
\begin{gather}
  \boldsymbol{\rho}=\sum_{N}{\cal{P}}_N\,\ket{\Psi_N}\bra{\Psi_N}.   
\end{gather}
Thus, each observable will lose contrast caused by this number uncertainty.
In this paper, we will establish an atom number filter for matter waves that
allows a number stabilization, i.\thinspace{}e., after passing each BEC
through the filter the number uncertainty is less than before.
\begin{figure}[h]
  \begin{center}
    \includegraphics[width=\columnwidth,angle=0]{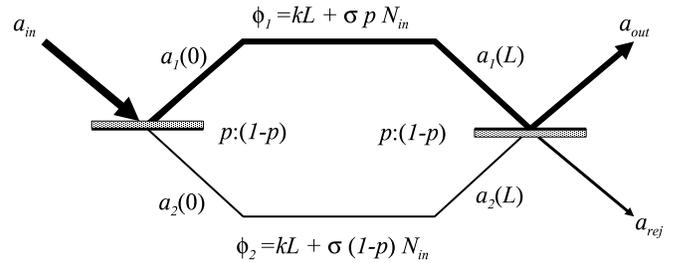}
    \caption{Setup for an asymmetric optical nonlinear interferometer
      with a propagation length $L$ and a splitting ratio $p:(1-p)$, $p \neq
      0.5$. Due to a Kerr-nonlinearity (susceptibility $\sigma$), one obtains
      a differential phase shift $\phi_{\rm NL}=\phi_1-\phi_2=\sigma (2p-1)
      N_{\rm in}$, proportional to an input photon intensity $N_{\rm in}$.  A
      subsequent self-interference of strong $a_1$ and weak field $a_2$
      stabilizes the output field intensity $N_{\rm out}=|a_{\rm out}|^2$ and
      diverts the noise to the rejection port $a_{\rm rej}$.}
    \label{optinterfer}
  \end{center}
\end{figure}

This can be achieved by using a nonlinear matter wave interferometer, cf.
Figs.~\ref{optinterfer} and \ref{interferometer}, which is in analogy to a
nonlinear fiber optics setup \cite{Schmitt1998}.  We will assume that the
condensate consists of atoms with two internal states. Due to the highly
asymmetric splitting, which is crucial in this setup, the condensate fraction
in one arm of the interferometer experiences a strong nonlinear phase
evolution, while the other part only undergoes a weak nonlinear phase shift.

The underlying physical mechanism of the suppression of number fluctuations is
based on the repulsive interaction amongst particles and has been used in the
context of spin squeezing \cite{ueda93,you03} or the Josephson effect
\cite{ivanov1999,leggett401,oberthaler06b}.  The ideas presented in here are
also related to the work of Poulsen and M\o{}lmer \cite{molmer02}, since their
approach combines ideas for light squeezing in a nonlinear optical
interferometer with the idea for spin-squeezing of two spatial, initially
identically occupied, condensate modes, generated via Bragg scattering.  Our
approach is different, as it explicitly requires a highly asymmetric splitting
$p\neq 0.5$ or would it disappear at all and it uses internal states of the
atom.

This paper is organized as follows: First, we briefly review the central idea
of amplitude stabilization of a nonlinear optical interferometer in
Sec.~\ref{optinterferometer}. Second, we introduce an equivalent model for a
bosonic matter wave in Sec.~\ref{matterinterferometer}. This is studied within
a mean-field picture to consider the effects of inhomogeneous traps as well as
a Josephson model of two quantized plane wave modes, which address the quantum
aspects and effects of finite particle numbers. Finally, conclusions are drawn
in Sec.~\ref{concl}.

\section{The principle of nonlinear interferometers in optical fibers}
\label{optinterferometer}
A very fundamental type of nonlinearity, which is present in many systems, is
the intensity-dependent phase shift. In photon optics it is due to the optical
Kerr effect \cite{agrawal} characterized by a susceptibility $\sigma$ and in
matter waves it is caused by interatomic atom forces measured by the s-wave
scattering length $a_s$. This leads to a self- or cross-phase modulation and
possibly to a self-trapping potential in the nonlinear Schr{\"o}dinger
equation of motion of wave-packets.

In the context of nonlinear interferometry \cite{Schmitt1998}, the intensity
filtering property is best in a highly asymmetric, highly transmissive
configuration, depicted in Fig.~\ref{optinterfer}. The interference of a
strong wave with a weak wave can eliminate a major fraction of the input noise
of $a_{in}$ in the output port $a_{out}$. A predominant part of the noise is
channeled to the rejection port $a_{rej}$, consuming a small fraction of the
input photon number.

\begin{figure}[h]
  \begin{center}
    \includegraphics[width=\columnwidth,angle=0]{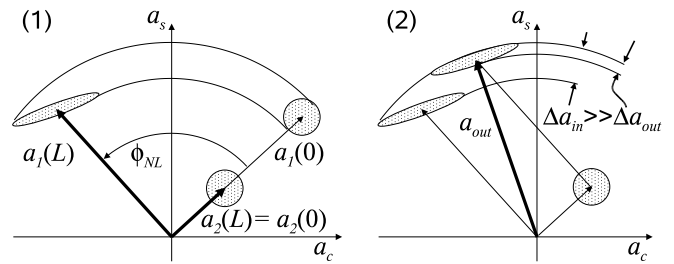}
    \caption{Schematic representation of the amplitude stabilization mechanism
      of $a_{\rm in}$ in an asymmetric, nonlinear interferometer in phase space
      with quadratures components $(a_{\rm c},a_{\rm s})$. After the first
      beam-splitter (1), the stronger field $a_1$ experiences a large
      nonlinear phase shift that translates amplitude fluctuations into
      correlated phase delay.  One can neglect the nonlinear phase shift of
      the weaker field $a_2$.  In a second beam-splitter (2), $a_2$ is
      coherently added to the phase-shifted field $a_1$ to cause output in
      $a_{\rm out}$. This stabilizes the output fluctuations well below the
      input level $\Delta a_{\rm in}\gg \Delta a_{\rm out}$, as indicated by
      the noise ellipses.}
    \label{phasespace}
  \end{center}
\end{figure}

Its two-step nonlinear self-stabilization mechanism is very simple and
visualized in Fig.~\ref{phasespace}.  After the first beam-splitter, the
asymmetric splitting of an input beam causes an intensity-dependent
differential phase shift between the two arms $a_1$ and $a_2$ of the
interferometer. The nonlinear phase shift transforms intensity amplitude
increase/reduction due to field fluctuations into a correlated phase
advance/delay.  Therefore, it causes a correlated phase spread relative to the
average nonlinear phase shift.  The second beam-splitter superposes both
interferometer beams coherently. It changes the quadrature angle of the field
amplitude relative to the noise distribution so that the amplitude-phase
correlation eliminates the amplitude noise to a large degree. For perfect
stabilization, the phase advance/delay of the stronger mode relative to the
weaker, quasi stationary, linearly propagating mode reduces/increases the
output transmission by the right amount to eliminate the input intensity
fluctuations.

It has been shown that the stabilization mechanism does not only eliminate
classical noise or works only with continuous-wave coherent light.  Instead,
this method is also applicable with broadband ultrashort solitons and in the
quantum regime of field fluctuations. The fiber-optic asymmetric Sagnac
interferometer has been used as a photon number filter for optical solitons
\cite{Schmitt1998, Sizmann1999}. Some of the best squeezing results have been
obtained with this set-up, which did not require any active stabilization.
The quantum noise reduction below the shot-noise (Poisson) limit has been
modeled by the quantum nonlinear Schr{\"o}dinger equation (NLSE). It is in
perfect agreement within the measurement uncertainty and stability has been
obtained.  Again, the noise reduction mechanism can be readily understood by
modeling the essentials of the asymmetric interferometer by linearized
fluctuations in a semiclassical approach where now the uncorrelated vacuum
fluctuations are entering through the unused input port of the interferometer
and where the associated phase is the soliton envelope phase. The
corresponding semiclassical picture is then well represented by
Fig.~\ref{phasespace}, where the noise ellipses are then the minimum
uncertainty regions.

Thus, the question arises, how the analogy of interference of bosonic fields
can be applied to matter waves. The analogy is not trivial, because the quanta
of the optical field and the massive bosons of the matter field are described
by different ensembles. Also, on a practical side, it can be asked how well
the asymmetric interferometer can function as a number filter for matter
waves.
\section{Modeling a nonlinear interferometer with bosonic matter-waves}
\label{matterinterferometer}
\begin{figure}[h]
\begin{center}
  \vspace{0.1cm} \includegraphics[width=\columnwidth,angle=0]{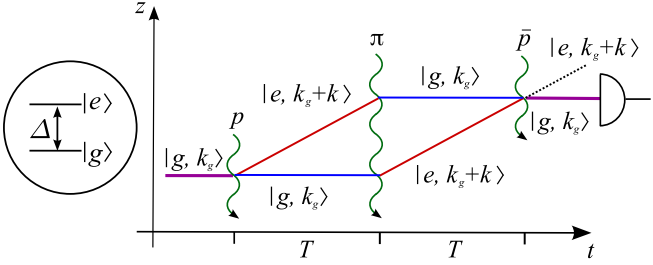}
  \caption{Setup for a matter-wave 
    interferometer. Absorption of a photon $k$ implements an asymmetric
    beam-splitter with splitting ratio $p:(1-p)$. After a free time evolution
    of duration $T$, an optional $\pi$-pulse inverts the populations. The
    second beam-splitter with splitting ratio $\bar{p}:(1-\bar{p})$ mixes the
    states for the final detection of one channel and the rejection of the
    other (comp.  Fig.~\ref{optinterfer}).}
\label{interferometer}
\end{center}
\end{figure}

Let us consider a trapped BEC consisting of two-level atoms labeled by
$\sigma=e,g$. The complete quantum states are then denoted by
$\ket{\sigma,\mathbf{k}_\sigma}$ with the internal state $\sigma$ and momentum
component $\mathbf{k}_\sigma$. The possibly time-dependent trapping potentials
for the two species are $V_\sigma(\mathbf{r},t)$.  They are identical
$V_e(\mathbf{r},t)=V_g(\mathbf{r},t)+\Delta$ up to a detuning $\Delta$.  The
two states are coupled by a classical traveling laser field
$\Omega(t)e^{i\mathbf{kr}}$ with the time-dependent Rabi frequency $\Omega(t)$
and the wave vector $\mathbf{k}$.  This configuration represents a
Ramsey-Bord\'e-interferometer (see Fig.~\ref{interferometer}) in the standard
setup of atom interferometry \cite{peters99}.  Initially, the BEC is prepared
in the $\ket{g}$-state and at an instant $t=0$ a $p$-pulse creates a
superposition of $\ket{g}$ and $\ket{e}$ with a splitting ratio of the
populations of $p:(1-p)$. After a time delay $T$, a further $\pi$-pulse gives
rise to an inversion of the populations. After another time interval $T$, a
final beam-splitter with splitting ratio $\bar{p}:(1-\bar{p})$ mixes the
populations again.

\subsection{Theoretical description}
In principle, the dynamics of this system is governed by the Schr\"odinger
equation $i\hbar\,\partial_t\ket{\psi(t)}=\hat{H}(t)\ket{\psi(t)}$ for a
many-particle state $\ket{\psi(t)}$ in Fock space and with the Hamiltonian
\begin{equation}
\label{ham}
\hat{H}(t)=\hat{H}_{\text{sp}}(t)+\hat{V}_{\text{d}}(t)+\hat{V}_{\text{p}},
\end{equation}
where
\begin{align*}
  \hat{H}_{\text{sp}}(t)=&
  \int\text{d}^3r\sum_{\sigma=e,g}\hat{a}^{\dagger}_{\sigma}(\mathbf{r})
  \left[-\frac{\hbar^2\nabla^2}{2m}+V_{\sigma}(\mathbf{r},t)\right]
  \hat{a}_{\sigma}(\mathbf{r}),\nonumber\\
  \hat{V}_{\text{d}}(t)=&\int\text{d}^3r\left[
    \hbar\,\Omega(t)e^{i{\mathbf{kr}}} \hat{a}^{\dagger}_{e} (\mathbf{r})
    \hat{a}^{\phantom \dagger}_{g}(\mathbf{r})
    +\text{h.c.}\right],\nonumber\\
  \hat{V}_{\text{p}}=&\tfrac{1}{2}\int\text{d}^3r\left[ g_{ee}
    \hat{a}^{\dagger}_{e}(\mathbf{r})\hat{a}^{\dagger}_{e}(\mathbf{r})
    \hat{a}^{\phantom \dagger}_{e}(\mathbf{r})
    \hat{a}^{\phantom \dagger}_{e}(\mathbf{r})\right.\nonumber\\
  &\left.+
    g_{gg}\hat{a}^{\dagger}_{g}(\mathbf{r})\hat{a}^{\dagger}_{g}(\mathbf{r})
    \hat{a}^{\phantom \dagger}_{g}(\mathbf{r})\hat{a}_{g}(\mathbf{r})\right.\nonumber\\
  &\left.+2 g_{eg}\hat{a}^{\dagger}_{e}(\mathbf{r})\hat{a}^{\dagger}_{g}
    (\mathbf{r})\hat{a}_{g}(\mathbf{r}) \hat{a}_{e}(\mathbf{r})\right].
\end{align*}
In there, we introduced bosonic field operators
\begin{gather}
[\hat{a}^{\phantom \dagger}_{\mu}(\mathbf{x}),
\hat{a}^{\dagger}_{\nu}(\mathbf{y})]=
\delta(\mathbf{x}-\mathbf{y})  \delta_{\mu \nu}
\end{gather}
and interatomic coupling constants $g_{\mu\nu}=4\pi\hbar^2a_{\mu\nu}/m$
between same and different species. $a_{\mu\nu}$ is the corresponding
scattering length and $m$ is the mass of a single particle.

In order to get physical insight into this problem, we consider two simplified
scenarios.  On the one hand, we study the classical field approximation in
Sec.~\ref{classfield}, where operators are replaced by complex amplitudes
$\hat{a}_{\sigma} \rightarrow \alpha_{\sigma}$, which yields the
Gross-Pitaevskii (GP) equation.  Thus, the statistical character of the
operator is neglected. Furthermore, we will consider a quasi one-dimensional,
cigar-shaped configuration with tight confinement in the radial direction
\cite{walser04}. The radial part can then be integrated out directly, which
results in modified coupling constants $g_{\mu\nu}\rightarrow \bar{g}_{\mu\nu}$. In
the following, we tacitly drop the bar. On the other hand, the operator
character is accounted for in the Josephson approximation of
Sec.~\ref{josephson}.  In there, we neglect spatial inhomogeneity and consider
only the behavior of two plane wave modes.

\subsection{Classical field approximation}
\label{classfield}
Within the classical field approximation, the corresponding scaled
quasi-one-dimensional GP equation reads
\begin{gather}
  \label{GP}
  i\partial_t
  \begin{pmatrix}
    \alpha_e(z,t)\\ \alpha_g(z,t)
  \end{pmatrix} =
  H_{\text{GP}}(t)
  \begin{pmatrix}
    \alpha_e\\ \alpha_g
  \end{pmatrix},\\
  H_{\text{GP}}(t)=  -\tfrac{1}{2}\partial_z^2+\begin{pmatrix}
  V_e(z,t)+v_e & \Omega(t)\,e^{ikz}\\
    \Omega^*(t)\,e^{-ikz} & V_g(z,t)+v_g
  \end{pmatrix},\nonumber
\end{gather}
where the mean-field energies are $v_\mu=g_{\mu\mu}n_\mu+g_{\mu\nu}n_\nu$ with
$\mu\neq \nu\in \{e,g\}$ and $n_\sigma=|\alpha_\sigma(z,t)|^2$. 

In the general case of time-dependent pulses and spatially inhomogeneous
traps, it is only possible to solve this equation numerically. Results of such
a calculation are presented in Sec.~\ref{inh}. However, if the system size is
large and time-dependent pulses happen on short times, then one can solve this
simplified situation analytically and gain qualitative understanding.

\subsubsection{A bulk BEC in the Raman-Nath approximation}
\label{bulk}
The beam-splitter is realized by a quasi-instantaneous ($\tau\ll
T,1/\Delta$) $p$-pulse in the form of a traveling laser wave with a Rabi
frequency $\Omega$.  Mathematically, this is described via the unitary
transformation
\begin{gather}
  \label{Ubarfirst}
  U_p= e^{-i\theta (e^{ikz}\sigma_+ +\text{h.c.})}
  =\begin{pmatrix}
    \cos \theta&-i\sin\theta e^{ikz}\\
    -i\sin \theta e^{-ikz}&\cos \theta
  \end{pmatrix},\nonumber\\
  \sigma_+=\begin{pmatrix} 0&1\\0&0\end{pmatrix},  \quad
  p=\sin^2\theta,\quad
\theta=\frac{\Omega\tau}{2}.
\end{gather}

In the following, we have propagated the mean-field state with flat potentials
$V_{g}(z,t)=0$, $V_{e}(z,t)=\Delta$ for a total time $2T$ in Eq.~(\ref{GP}).
The free nonlinear evolution was interrupted by the interferometer pulse
sequence $p-\pi-\bar{p}$ depicted in Fig.~\ref{interferometer}.  We assume
that all the population $n$ is initially in the ground state component
$\{\alpha_e(0),\alpha_g(0)\}=\{0,\sqrt{n}\,e^{ik_g z}\}$, moving with a
generic momentum $k_g$.  The general solution for the free propagation is
obtained easily by making the plane-wave ansatz
\begin{gather}
\{\alpha_e(z,t),\alpha_g(z,t)\}=
\{\alpha_e e^{i(k_e z-\phi_e)},\alpha_g e^{i(k_g z-\phi_g)}\},
\end{gather}  
with time-dependent phases $\phi_\sigma(t)$ and a recoil-shifted momentum
$k_e=k_g+k$.  We are interested in the particle density of the output channel
$n_\sigma^{\text{out}} =n_\sigma(2T)=|\alpha_\sigma(2T)|^2$ of the
interferometer. After simple algebra, one finds for the transmitted channels
\begin{gather}
  \label{neg}
  n_{e}^{\text{out}}= n
  \left( \xi-\gamma\cos\left[2T n\delta_2 (p-\tfrac{1}{2})\right] \right),\\
  \xi=p\bar{p}+(1-p)(1-\bar{p}),\quad
  \gamma=2\sqrt{p\bar{p}(1-p)(1-\bar{p})},\nonumber
\end{gather}
with $\delta_2=g_{ee}-2g_{eg}+g_{gg}$ a central difference of scattering
lengths. The population in the other channel
$n_{g}^{\text{out}}=n-n_{e}^{\text{out}}$ follows from number conservation.

A simplified but less efficient form of the nonlinear interferometer is found
from a symmetric mixing $\bar{p}=\tfrac{1}{2}$ at the final output beam-splitter
\begin{equation}
 \label{negsim}
  n_{e}^{\text{out}}=
  n\left(\tfrac{1}{2}-\sqrt{p(1-p)}
    \cos\left[ 2T n\delta_2(p-\tfrac{1}{2})\right]\right).
\end{equation}
The most salient features of Eqs.~(\ref{neg}) and (\ref{negsim}) are the
absence of linear phase shifts due to the intermediate $\pi$-pulse and the
nonlinear phase shift. It is proportional to the total interaction time $2T$,
to the density $n$, the central difference $\delta_2$ of scattering lengths,
as well as the asymmetry $p\neq 0.5$ of initial beam-splitting.
The oscillatory response of the interferometer with respect to a varying input
particle number $n$ stabilizes the output particle number $n^{\text{out}}_e$,
if operated in the vicinity of an extremum. This suppression of number
fluctuations represents a nonlinear number filter for matter waves and is
depicted in Fig.~\ref{twomodepicture}.  It is also important to note that a
symmetric splitting $p=\tfrac{1}{2}$, or vanishing central difference
$\delta_2=0$ of scattering lengths lead to no effect at all.

Alternatively, if one considers a simplified interferometer setup in
Fig.~\ref{interferometer}, i.\thinspace{}e., without the intermediate
$\pi$-pulse and chooses a symmetric final beam-splitting $\bar{p}=\tfrac{1}{2}$,
one obtains
\begin{equation}
  \label{sol_2pulses}
  n_{e}^{\text{out}}= n\left(\tfrac{1}{2}
    +\sqrt{p(1-p)}\cos[\{\Delta_D+(\delta_1+p\delta_2)n\}T]\right),
\end{equation}
with a total propagation time $T$. This interferometer is sensitive
to single particle phase shifts, here in particular to the Doppler-shifted
detuning $\Delta_D=\Delta+k_e^2/2$ and also the difference of scattering
lengths $\delta_1=g_{eg}-g_{gg}$. For example, this setup is used for
atom gravitometry \cite{peters99}, but it is also more susceptible to
experimental noise (frequency jigger), which deteriorates visibility.
Nevertheless, it is favorable for $^{87}$Rb BEC's, where the coupling
constants are $g_{ee}:g_{eg}:g_{gg}=1.03:1:0.97$, hence $\delta_2$ vanishes.
In contrast, there is an effect in the latter setup, since the difference
$\delta_1$ is finite.

\subsubsection{An inhomogeneous BEC in a square well trap}
\label{inh}

In order to study the reduction of the filter performance caused by the
inherent inhomogeneity of trapped atomic BECs, we have chosen a square-well
potential
\begin{equation}
V_g(z,t)=\left\{  
  \begin{matrix}
    V_g<0,&t<0,& |z|<L\\
    0,& t\geq 0
  \end{matrix}\right. ,
\end{equation}
to account for the initial inhomogeneity of the ground-state.  After the first
$p$-pulse, the trap is switched off permanently. The excited state potential
is identical, but shifted in energy by the detuning from the laser,
i.\thinspace{}e., $V_e(z,t)=V_g(z,t)+\Delta$.
\begin{figure}[h]
\begin{center}
  \includegraphics[width=\columnwidth,angle=0]{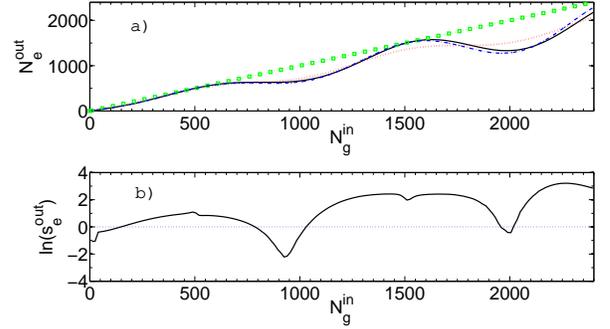}
\caption{Response of a highly asymmetric $p=\bar{p}=0.9$, nonlinear 
  $p-\pi-\bar{p}$ matter-wave interferometer.  a) Outgoing particles in the
  e-channel $N_{e}^{\text{out}}$ versus the incoming number
  $N_{g}^{\text{in}}$ in the g-channel: homogeneous mean-field
  (dashed-dotted), inhomogeneous mean-field (dotted), and two-mode
  approximation (solid). The line marked with $\square$ is the trivial
  response of a $0-\pi-0$ interferometer, i.\thinspace{}e.,
  $N_{e}^{\text{out}}=N_g^{\text{in}}$ and can be used to assess the loss of
  particles in the unobserved channel. b) Corresponding normalized number
  fluctuations $s_e^{\text{out}}$ in a semilogarithmic representation.  The
  optimal working point for the interferometer as a number-filter device is
  $N_{g}^{\text{in}}=940$ (see Fig.~\ref{prob}). There, one finds a strong,
  sub-shot noise suppression of number fluctuations in the output channel
  $s_e^{\text{out}}\ll 1$ .}
\label{twomodepicture}
\end{center}
\end{figure}

In the numerical simulations for the homogeneous and inhomogeneous
condensates, we have used generic parameters for a quasi-1d elongated
$^{87}$Rb BEC with an atomic mass $m=87\,\text{amu}$. In a prolate harmonic
oscillator with a trapping frequency $\omega_z=4$\,1/s, the basic length unit
would be $a_{\text{ho}}=\sqrt{\hbar/m \omega_z}=13.2\,\mu\text{m}$.  We have
used a multiple of this scale for the length of the square-well trap
$L=132\,\mu\text{m}\,(10)$. Potential depths
$V_g=-\hbar\,60\,\text{Hz}\,(-15)$ and detunings $\Delta=0$ are measured in
natural energy units $\hbar \omega_z$.  In a copropagating Raman laser
configuration, one can have a vanishing momentum transfer $k=0$ and we assumed
that the condensate is initially at rest $k_g=0$.  Short laser pulses were
used such that the beam-splitters were highly asymmetric $p=\bar{p}=0.9$ and
the propagation time between the pulses was $T=20\,\text{ms}$.  We have
deliberately modified the s-wave scattering lengths for $^{87}$Rb slightly to
obtain dimensionless quasi-1d coupling constants $g_{ee}=0.034$,
$g_{eg}=0.10$, and $g_{gg}=0.068$. Thus, the superior $p-\pi-\bar{p}$
interferometer scheme remains applicable. In principle, this can be achieved
via Feshbach resonances or using other elements like $^{85}$Rb or $^{23}$Na
altogether.

For this situation, we have numerically solved the time-dependent inhomogeneous
GP Eq.~(\ref{GP}) and find a similar behavior as for the homogeneous limit in
Fig.~\ref{twomodepicture}. As expected, we obtain a number stabilization of
the nonlinear filter, but at a slightly reduced performance due to the
inhomogeneous averaging.  We have also verified that the best number
stabilization is achieved for highly asymmetric splitting, that the effect
does not occur for equal scattering lengths, or linear matter-wave
interferometers at all. Moreover, the assumed square-well trap is not a
peculiarity in this context, as we have tried a harmonic oscillator trap
and find qualitatively similar results.

\subsection{Quantum mechanical two-mode approximation}
\label{josephson}
In the classical field approximation for a homogeneous bulk system of
Sec.~\ref{bulk}, we have examined the static number filter response of the
interferometer. The macroscopically occupied amplitudes were described like
two coherently coupled nonlinear oscillators. They exhibited a noise
suppression that is analogously used in many other physical systems ranging
from electrical circuits to coupled Josephson junctions
\cite{barone82,leggett91,goldobin05}.

In order to probe the quantum aspects of such a system, we will assume now
that the atomic fields $\hat{a}_\sigma(z)$ are dominated by two plane-wave
modes labeled with bosonic field amplitudes $\hat{e}$ and $\hat{g}$
\begin{gather}
  \hat{a}_e(z)=\hat{e}\,e^{ik_ez}+\delta \hat{a}_e, \quad
  \hat{a}_g(z)=\hat{g}\,e^{ik_gz}+\delta \hat{a}_g.
\end{gather}
Their residual coupling to other modes $\delta \hat{a}_\sigma$, is small at
the relevant time scales and will be disregarded altogether.

Within this Josephson approximation and
in the quasi-1d configuration, we can simplify the Hamiltonian of
Eq.~(\ref{ham}) further to $\hat{H}\approx \hat{H}_J$ with
\begin{gather}
\begin{aligned}
  \hat{H}_{J}&= \Delta_D\hat{n}_{e}
  +\frac{k_g^2}{2}\hat{n}_{g}
  +\Omega(t)\hat{e}^{\dagger}\hat{g}+
\Omega^\ast(t)\hat{g}^{\dagger}\hat{e}+\\
  &\tfrac{1}{2} g_{ee}\hat{n}_{e}(\hat{n}_{e}-1)
  +\tfrac{1}{2} g _{gg}\hat{n}_{g}(\hat{n}_{g}-1)
  +g_{eg}\hat{n}_{e}\hat{n}_{g},
\end{aligned}  
\end{gather}
where $\hat{n}_{\sigma}=\{ \hat{e}^{\dagger}\hat{e}^{\phantom \dagger},
\hat{g}^{\dagger} \hat{g}^{\phantom \dagger}\}$ denotes the particle number
operator for the two modes and $\Delta_D=\Delta+k_e^2/2$ is the
Doppler-shifted detuning as in Eq.~(\ref{sol_2pulses}).  The consistency of
the limit can be checked quickly by replacing the mode operators again by
complex numbers, just to recover Eq.~(\ref{GP}).

Clearly, the Josephson Hamiltonian conserves the particle number,
\begin{gather}
  \hat{N}=\hat{n}_e+\hat{n}_g, \quad [\hat{H}_J,\hat{N}]=0.
\end{gather}
Thus, a general state in the $N$-particle sector of Fock space is given by a
superposition of the states $\ket{n_e,n_g}$ with $N=n_e+n_g$
\begin{gather}
  \ket{\psi_N(t)}=\sum_{n=0}^{N} \psi_N^n(t)\ket{N-n,n}.
\end{gather}
The time evolution of such a state leads to a simple one-dimensional
difference equation for the time-dependent amplitudes
\begin{align}
  \label{difference}
  i\dot{\psi}_N^n(t)=&w^n \psi_N^n+q^n(t) \psi^{n-1}_N+{q^{n+1}}^\ast(t)
  \psi^{n+1}_N,\\
  w^n=&\Delta_D(N-n)+\frac{k_g^2}{2}n+\tfrac{1}{2}g_{gg}n(n-1)+\\
  &\tfrac{1}{2}g_{ee}(N-n)(N-n-1)+g_{eg}\,n(N-n),\nonumber\\
  q^n(t)=&\Omega(t)\sqrt{n(N-n+1)},
\end{align}
which can be solved easily on a computer or approximated analytically
\cite{braun93}.

We are now in a position to discuss the number stabilization scheme on a full
quantum mechanical level.  Given that a single realization of a BEC had a
well-defined number $N$, initially with all atoms in the ground state
component $\ket{\psi_N(t=0)}=\ket{0,n_g=N}$, then there is obviously no need
for an extra number filtering device, as it is sharply defined by perfect
preparation.  However, the experimental reality usually is plagued with
technical imperfections, day-to-day variations or finite temperatures
ensembles. Thus, one should include the possible number uncertainty in a
statistical description and use the density matrix
\begin{gather}
  \boldsymbol{\rho}(t)=\sum_{N=0}^{\infty}\mathcal{P}_N\,\ket{\psi_N(t)}
  \bra{\psi_N(t)},
\end{gather}
where the distribution $\cal{P}_N$ accounts for the lack of information
\cite{schleich01}.  As usual, we obtain averages of observable with respect to
the mixed state by a trace $\langle \ldots \rangle=\text{Tr}\{\ldots
\boldsymbol{\rho}(t) \}$ over all the Fock space.
\begin{figure}[h]
\begin{center}
\includegraphics[width=\columnwidth,angle=0]{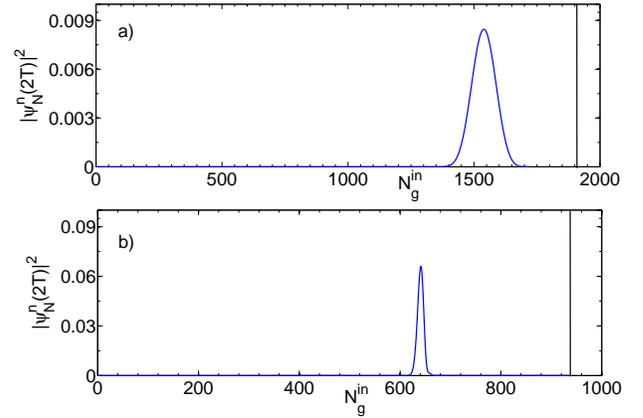}
\caption{
  Evolution of the particle number distribution $|\psi_N^n(2T)|^2$ from an
  initially pure Fock state (solid vertical line) to a broadened distribution,
  exiting in the interferometer port e. In subplot a), we used
  $\ket{0,N=1720}$, which leads to a large number dispersion. In subplot b) we
  used the particle number $\ket{0,N=940}$ that defines the optimal working
  point of the interferometer in Fig.~\ref{twomodepicture}.}
\label{prob}
\end{center}
\end{figure}

The particle output in the e-channel of the interferometric number filter and
its uncertainty are measured by averages, variances and volatility
\begin{gather}
  \langle \hat{n}_e(t)\rangle=\sum_{N=0}^{\infty}{\cal P}_N
  \bra{\psi_N(t)}\hat{n}_e\ket{\psi_N(t)}=\nonumber\\
  \sum_{N=0}^{\infty}{\cal P}_N \sum_{n=0}^N (N-n) |\psi_N^n(t)|^2,\\
  \sigma_e^2(t)=\langle \hat{n}_e^2(t)\rangle -\langle \hat{n}_e(t)
  \rangle^2,\quad s_e^{\text{out}}=\frac{\sigma_e(2T)}{\langle
    \hat{n}_e(2T)\rangle}.
\end{gather}

We have now evaluated the time-dependent difference Schr\"odinger equation
Eq.~(\ref{difference}) for the $p-\pi-\bar{p}$ interferometer for a range of
initial particle numbers $0\leq N \leq 2500$ in the BEC. In particular, we
have even relaxed the Raman-Nath approximation for the $p-\pi-\bar{p}$
beam-splitter sequence and used real rectangular pulses with duration
$\tau=1.31$ ms$\ll T$ and total pulse areas $\Omega \tau $ such that
$p=\bar{p}=0.9$, as before.  All other parameters were as in Sec.~{\ref{inh}}.
The results of this quantum mechanical simulation, in particular the mean
output number $\langle n_e(2T)\rangle$ and normalized number uncertainty
$s^{\text{out}}_e$ are depicted in Fig.~\ref{twomodepicture}. They agree well
with the homogeneous and inhomogeneous mean-field calculations.

In Fig.~\ref{prob}, we show two special wavefunctions: one that corresponds to
the optimal input number $N=940$ and one with $N=1720$ that exhibits a large
number dispersion after passing through the interferometer. The Gaussian
nature of the final wavefunction can be explained from a semiclassical
analysis of the difference equation \cite{braun93}. Both wave functions can be
identified also clearly as extrema of the response function in
Fig.~\ref{twomodepicture}. From these two extreme examples of the evolution of
pure initial Fock states it is obvious that a further convolution with a
non-ideally prepared ensemble, for example $\mathcal{P}_N \sim
\exp{[-(N-\bar{N})^2/\bar{\sigma}]}$, will exhibit a similar dispersion
response: a suppression of number fluctuation in the output channel if the
optimal working point of the filter coincides with the most likely particle
number of the input ensemble.

\section{Conclusion}
\label{concl}
In conclusion, we have analyzed the performance of a nonlinear number filter
for matter waves. This addresses the problem of technical shot-to-shot
variation of particle number in current Bose-Einstein condensate experiments.
A highly asymmetric $p-\pi-\bar{p}$ beam-splitter sequence is required to
achieve optimal filtering performance. In the case of symmetric splitting or
the absence of nonlinearity, no filtering is seen at all.  This method is in
direct analogy to a nonlinear fiber optics setup studied in
\cite{Kitagawa1986,Schmitt1998,Sizmann1999}.  In detail, we have analyzed the
situation for an homogeneous system and an inhomogeneous trapped gas within
the Gross-Pitaevskii mean-field theory, as well as a quantum mechanical
Josephson model, which addresses complementary aspects and agrees well
otherwise.

We thank E. Kajari for fruitful discussions. G.~N. and R.~W.  acknowledge
gratefully financial support by the Deutsches Zentrum f\"ur Luft- und
Raumfahrt (50 WM 0346), the Landesstifung Baden-W\"urttemberg
(AKZ0904Atom14), as well as the ESA programm SAI (AO-2004-64).
\bibliography{bec,MyPublications,ref_droptower}

\begin{thebibliography}{56}
\expandafter\ifx\csname natexlab\endcsname\relax\def\natexlab#1{#1}\fi
\expandafter\ifx\csname bibnamefont\endcsname\relax
  \def\bibnamefont#1{#1}\fi
\expandafter\ifx\csname bibfnamefont\endcsname\relax
  \def\bibfnamefont#1{#1}\fi
\expandafter\ifx\csname citenamefont\endcsname\relax
  \def\citenamefont#1{#1}\fi
\expandafter\ifx\csname url\endcsname\relax
  \def\url#1{\texttt{#1}}\fi
\expandafter\ifx\csname urlprefix\endcsname\relax\def\urlprefix{URL }\fi
\providecommand{\bibinfo}[2]{#2}
\providecommand{\eprint}[2][]{\url{#2}}

\bibitem[{\citenamefont{Agrawal}(2001)}]{agrawal}
\bibinfo{author}{\bibfnamefont{G.}~\bibnamefont{Agrawal}},
  \emph{\bibinfo{title}{Nonlinear Fiber Optics}} (\bibinfo{publisher}{Academic
  Press}, \bibinfo{year}{2001}).

\bibitem[{\citenamefont{Mitra and Stark}(2001)}]{Mitra2001}
\bibinfo{author}{\bibfnamefont{P.}~\bibnamefont{Mitra}} \bibnamefont{and}
  \bibinfo{author}{\bibfnamefont{J.}~\bibnamefont{Stark}},
  \bibinfo{journal}{Nature} \textbf{\bibinfo{volume}{411}},
  \bibinfo{pages}{1027} (\bibinfo{year}{2001}).

\bibitem[{\citenamefont{Jaekel and Reynaud}(1990)}]{Jaekel1990}
\bibinfo{author}{\bibfnamefont{M.}~\bibnamefont{Jaekel}} \bibnamefont{and}
  \bibinfo{author}{\bibfnamefont{S.}~\bibnamefont{Reynaud}},
  \bibinfo{journal}{Europhys. Lett.} \textbf{\bibinfo{volume}{13}},
  \bibinfo{pages}{301} (\bibinfo{year}{1990}).

\bibitem[{\citenamefont{Ritze and Bandilla}(1979)}]{Ritze1979}
\bibinfo{author}{\bibfnamefont{H.}~\bibnamefont{Ritze}} \bibnamefont{and}
  \bibinfo{author}{\bibfnamefont{A.}~\bibnamefont{Bandilla}},
  \bibinfo{journal}{Opt. Comm.} \textbf{\bibinfo{volume}{29}},
  \bibinfo{pages}{126} (\bibinfo{year}{1979}).

\bibitem[{\citenamefont{Kaplan and Meystre}(1981)}]{kaplan81}
\bibinfo{author}{\bibfnamefont{A.~E.} \bibnamefont{Kaplan}} \bibnamefont{and}
  \bibinfo{author}{\bibfnamefont{P.}~\bibnamefont{Meystre}},
  \bibinfo{journal}{Opt. Lett.} \textbf{\bibinfo{volume}{6}},
  \bibinfo{pages}{590} (\bibinfo{year}{1981}).

\bibitem[{\citenamefont{Kitagawa and Yamamoto}(1986)}]{Kitagawa1986}
\bibinfo{author}{\bibfnamefont{M.}~\bibnamefont{Kitagawa}} \bibnamefont{and}
  \bibinfo{author}{\bibfnamefont{Y.}~\bibnamefont{Yamamoto}},
  \bibinfo{journal}{Phys.~Rev.~A} \textbf{\bibinfo{volume}{34}},
  \bibinfo{pages}{3974} (\bibinfo{year}{1986}).

\bibitem[{\citenamefont{Schmitt et~al.}(1998)\citenamefont{Schmitt, Ficker,
  Wolff, K\"o{}nig, Sizmann, and Leuchs}}]{Schmitt1998}
\bibinfo{author}{\bibfnamefont{S.}~\bibnamefont{Schmitt}},
  \bibinfo{author}{\bibfnamefont{J.}~\bibnamefont{Ficker}},
  \bibinfo{author}{\bibfnamefont{M.}~\bibnamefont{Wolff}},
  \bibinfo{author}{\bibfnamefont{F.}~\bibnamefont{K\"o{}nig}},
  \bibinfo{author}{\bibfnamefont{A.}~\bibnamefont{Sizmann}}, \bibnamefont{and}
  \bibinfo{author}{\bibfnamefont{G.}~\bibnamefont{Leuchs}},
  \bibinfo{journal}{Phys.~Rev.~Lett.} \textbf{\bibinfo{volume}{81}},
  \bibinfo{pages}{2446} (\bibinfo{year}{1998}).

\bibitem[{\citenamefont{Werner}(1998)}]{Werner1998b}
\bibinfo{author}{\bibfnamefont{M.}~\bibnamefont{Werner}},
  \bibinfo{journal}{Phys.~Rev.~Lett.} \textbf{\bibinfo{volume}{81}},
  \bibinfo{pages}{4132} (\bibinfo{year}{1998}).

\bibitem[{\citenamefont{Doran and Wood}(1988)}]{wood88}
\bibinfo{author}{\bibfnamefont{N.~J.} \bibnamefont{Doran}} \bibnamefont{and}
  \bibinfo{author}{\bibfnamefont{D.}~\bibnamefont{Wood}},
  \bibinfo{journal}{Opt. Lett.} \textbf{\bibinfo{volume}{13}},
  \bibinfo{pages}{56} (\bibinfo{year}{1988}).

\bibitem[{\citenamefont{Sizmann and Leuchs}(1999)}]{Sizmann1999}
\bibinfo{author}{\bibfnamefont{A.}~\bibnamefont{Sizmann}} \bibnamefont{and}
  \bibinfo{author}{\bibfnamefont{G.}~\bibnamefont{Leuchs}},
  \bibinfo{journal}{Prog. Opt.} \textbf{\bibinfo{volume}{39}},
  \bibinfo{pages}{373} (\bibinfo{year}{1999}).

\bibitem[{\citenamefont{Levandovsky et~al.}(1999)\citenamefont{Levandovsky,
  Vasilyev, and Kumar}}]{Levandovsky1999}
\bibinfo{author}{\bibfnamefont{D.}~\bibnamefont{Levandovsky}},
  \bibinfo{author}{\bibfnamefont{M.}~\bibnamefont{Vasilyev}}, \bibnamefont{and}
  \bibinfo{author}{\bibfnamefont{P.}~\bibnamefont{Kumar}},
  \bibinfo{journal}{Opt. Lett.} \textbf{\bibinfo{volume}{24}},
  \bibinfo{pages}{89} (\bibinfo{year}{1999}).

\bibitem[{\citenamefont{Bachor and Ralph}(2003)}]{Bachor2003}
\bibinfo{author}{\bibfnamefont{H.}~\bibnamefont{Bachor}} \bibnamefont{and}
  \bibinfo{author}{\bibfnamefont{T.}~\bibnamefont{Ralph}},
  \emph{\bibinfo{title}{A Guide to Experiments in Quantum Optics}}
  (\bibinfo{publisher}{Wiley-VCH}, \bibinfo{year}{2003}).

\bibitem[{\citenamefont{Schleich}(2001)}]{schleich01}
\bibinfo{author}{\bibfnamefont{W.~P.} \bibnamefont{Schleich}},
  \emph{\bibinfo{title}{Quantum Optics in Phase Space}}
  (\bibinfo{publisher}{Wiley-VCH}, \bibinfo{address}{Berlin, Germany},
  \bibinfo{year}{2001}).

\bibitem[{\citenamefont{M\o{}lmer}(2005)}]{moelmer05}
\bibinfo{author}{\bibfnamefont{K.}~\bibnamefont{M\o{}lmer}},
  \bibinfo{journal}{N. J. P.} \textbf{\bibinfo{volume}{5}}, \bibinfo{pages}{55}
  (\bibinfo{year}{2005}).

\bibitem[{\citenamefont{Yasuda and Shimizu}(1996)}]{shimitsu96}
\bibinfo{author}{\bibfnamefont{M.}~\bibnamefont{Yasuda}} \bibnamefont{and}
  \bibinfo{author}{\bibfnamefont{F.}~\bibnamefont{Shimizu}},
  \bibinfo{journal}{Phys. Rev. Lett.} \textbf{\bibinfo{volume}{77}},
  \bibinfo{pages}{3090} (\bibinfo{year}{1996}).

\bibitem[{\citenamefont{Andrews et~al.}(1997)\citenamefont{Andrews, Townsend,
  Miesner, Durfee, Kurn, and Ketterle}}]{ketterleinterference}
\bibinfo{author}{\bibfnamefont{M.}~\bibnamefont{Andrews}},
  \bibinfo{author}{\bibfnamefont{C.}~\bibnamefont{Townsend}},
  \bibinfo{author}{\bibfnamefont{H.-J.} \bibnamefont{Miesner}},
  \bibinfo{author}{\bibfnamefont{D.}~\bibnamefont{Durfee}},
  \bibinfo{author}{\bibfnamefont{D.}~\bibnamefont{Kurn}}, \bibnamefont{and}
  \bibinfo{author}{\bibfnamefont{W.}~\bibnamefont{Ketterle}},
  \bibinfo{journal}{Science} \textbf{\bibinfo{volume}{275}},
  \bibinfo{pages}{637} (\bibinfo{year}{1997}).

\bibitem[{\citenamefont{Saubam\'ea et~al.}(1997)\citenamefont{Saubam\'ea,
  Hijmans, Kulin, Rasel, Peik, Leduc, and Cohen-Tannoudji}}]{CohenTannoudji97}
\bibinfo{author}{\bibfnamefont{B.}~\bibnamefont{Saubam\'ea}},
  \bibinfo{author}{\bibfnamefont{T.~W.} \bibnamefont{Hijmans}},
  \bibinfo{author}{\bibfnamefont{S.}~\bibnamefont{Kulin}},
  \bibinfo{author}{\bibfnamefont{E.}~\bibnamefont{Rasel}},
  \bibinfo{author}{\bibfnamefont{E.}~\bibnamefont{Peik}},
  \bibinfo{author}{\bibfnamefont{M.}~\bibnamefont{Leduc}}, \bibnamefont{and}
  \bibinfo{author}{\bibfnamefont{C.}~\bibnamefont{Cohen-Tannoudji}},
  \bibinfo{journal}{Phys. Rev. Lett.} \textbf{\bibinfo{volume}{79}},
  \bibinfo{pages}{3146} (\bibinfo{year}{1997}).

\bibitem[{\citenamefont{Bloch et~al.}(2000)\citenamefont{Bloch, H{\"a}nsch, and
  Esslinger}}]{esslinger02}
\bibinfo{author}{\bibfnamefont{I.}~\bibnamefont{Bloch}},
  \bibinfo{author}{\bibfnamefont{T.}~\bibnamefont{H{\"a}nsch}},
  \bibnamefont{and}
  \bibinfo{author}{\bibfnamefont{T.}~\bibnamefont{Esslinger}},
  \bibinfo{journal}{Nature} \textbf{\bibinfo{volume}{403}},
  \bibinfo{pages}{166} (\bibinfo{year}{2000}).

\bibitem[{\citenamefont{Orzel et~al.}(2001)\citenamefont{Orzel, Tuchman,
  Fenselau, and M.~Yasuda}}]{Orzel01}
\bibinfo{author}{\bibfnamefont{C.}~\bibnamefont{Orzel}},
  \bibinfo{author}{\bibfnamefont{A.}~\bibnamefont{Tuchman}},
  \bibinfo{author}{\bibfnamefont{M.}~\bibnamefont{Fenselau}}, \bibnamefont{and}
  \bibinfo{author}{\bibfnamefont{M.~K.} \bibnamefont{M.~Yasuda}},
  \bibinfo{journal}{Science} \textbf{\bibinfo{volume}{291}},
  \bibinfo{pages}{2386} (\bibinfo{year}{2001}).

\bibitem[{\citenamefont{Schellekens et~al.}(2005)\citenamefont{Schellekens,
  Hoppeler, Perrin, Gomes, Boiron, Aspect, and Westbrook}}]{westbrook05}
\bibinfo{author}{\bibfnamefont{M.}~\bibnamefont{Schellekens}},
  \bibinfo{author}{\bibfnamefont{R.}~\bibnamefont{Hoppeler}},
  \bibinfo{author}{\bibfnamefont{A.}~\bibnamefont{Perrin}},
  \bibinfo{author}{\bibfnamefont{J.~V.} \bibnamefont{Gomes}},
  \bibinfo{author}{\bibfnamefont{D.}~\bibnamefont{Boiron}},
  \bibinfo{author}{\bibfnamefont{A.}~\bibnamefont{Aspect}}, \bibnamefont{and}
  \bibinfo{author}{\bibfnamefont{C.}~\bibnamefont{Westbrook}},
  \bibinfo{journal}{Science} \textbf{\bibinfo{volume}{310}},
  \bibinfo{pages}{648} (\bibinfo{year}{2005}).

\bibitem[{\citenamefont{Chuu et~al.}(2005)\citenamefont{Chuu, Schreck, Meyrath,
  Hanssen, Price, and Raizen}}]{Raizen2005a}
\bibinfo{author}{\bibfnamefont{C.-S.} \bibnamefont{Chuu}},
  \bibinfo{author}{\bibfnamefont{F.}~\bibnamefont{Schreck}},
  \bibinfo{author}{\bibfnamefont{T.~P.} \bibnamefont{Meyrath}},
  \bibinfo{author}{\bibfnamefont{J.~L.} \bibnamefont{Hanssen}},
  \bibinfo{author}{\bibfnamefont{G.~N.} \bibnamefont{Price}}, \bibnamefont{and}
  \bibinfo{author}{\bibfnamefont{M.~G.} \bibnamefont{Raizen}},
  \bibinfo{journal}{Phys.~Rev.~Lett.} \textbf{\bibinfo{volume}{95}},
  \bibinfo{eid}{260403} (\bibinfo{year}{2005}).

\bibitem[{\citenamefont{Jeltes et~al.}(2007)\citenamefont{Jeltes, McNamara,
  Hogervorst, Vassen, Krachmalnicoff, Schellekens, Perrin, Chang, Boiron,
  Aspect et~al.}}]{westbrook07}
\bibinfo{author}{\bibfnamefont{T.}~\bibnamefont{Jeltes}},
  \bibinfo{author}{\bibfnamefont{J.}~\bibnamefont{McNamara}},
  \bibinfo{author}{\bibfnamefont{W.}~\bibnamefont{Hogervorst}},
  \bibinfo{author}{\bibfnamefont{W.}~\bibnamefont{Vassen}},
  \bibinfo{author}{\bibfnamefont{V.}~\bibnamefont{Krachmalnicoff}},
  \bibinfo{author}{\bibfnamefont{M.}~\bibnamefont{Schellekens}},
  \bibinfo{author}{\bibfnamefont{A.}~\bibnamefont{Perrin}},
  \bibinfo{author}{\bibfnamefont{H.}~\bibnamefont{Chang}},
  \bibinfo{author}{\bibfnamefont{D.}~\bibnamefont{Boiron}},
  \bibinfo{author}{\bibfnamefont{A.}~\bibnamefont{Aspect}},
  \bibnamefont{et~al.}, \bibinfo{journal}{Nature}
  \textbf{\bibinfo{volume}{445}}, \bibinfo{pages}{402} (\bibinfo{year}{2007}).

\bibitem[{\citenamefont{Fort{\'a}gh and Zimmermann}(2007)}]{fortagh07}
\bibinfo{author}{\bibfnamefont{J.}~\bibnamefont{Fort{\'a}gh}} \bibnamefont{and}
  \bibinfo{author}{\bibfnamefont{C.}~\bibnamefont{Zimmermann}},
  \bibinfo{journal}{Rev. Mod. Phys.} \textbf{\bibinfo{volume}{79}},
  \bibinfo{pages}{235} (\bibinfo{year}{2007}).

\bibitem[{\citenamefont{Jaksch et~al.}(1998)\citenamefont{Jaksch, Bruder,
  Cirac, Gardiner, and Zoller}}]{jaksch98}
\bibinfo{author}{\bibfnamefont{D.}~\bibnamefont{Jaksch}},
  \bibinfo{author}{\bibfnamefont{C.}~\bibnamefont{Bruder}},
  \bibinfo{author}{\bibfnamefont{J.}~\bibnamefont{Cirac}},
  \bibinfo{author}{\bibfnamefont{C.}~\bibnamefont{Gardiner}}, \bibnamefont{and}
  \bibinfo{author}{\bibfnamefont{P.}~\bibnamefont{Zoller}},
  \bibinfo{journal}{Phys.~Rev.~Lett.} \textbf{\bibinfo{volume}{81}},
  \bibinfo{pages}{3108} (\bibinfo{year}{1998}).

\bibitem[{\citenamefont{G{\"o}rlitz et~al.}(2001)\citenamefont{G{\"o}rlitz,
  Vogels, Leanhardt, Raman, ad~J.~Abo-Shaeer, Chikkatur, Gupta, Inouye,
  Rosenband, and Ketterle}}]{goerlitz01}
\bibinfo{author}{\bibfnamefont{A.}~\bibnamefont{G{\"o}rlitz}},
  \bibinfo{author}{\bibfnamefont{J.}~\bibnamefont{Vogels}},
  \bibinfo{author}{\bibfnamefont{A.}~\bibnamefont{Leanhardt}},
  \bibinfo{author}{\bibfnamefont{C.}~\bibnamefont{Raman}},
  \bibinfo{author}{\bibfnamefont{T.~G.} \bibnamefont{ad~J.~Abo-Shaeer}},
  \bibinfo{author}{\bibfnamefont{A.}~\bibnamefont{Chikkatur}},
  \bibinfo{author}{\bibfnamefont{S.}~\bibnamefont{Gupta}},
  \bibinfo{author}{\bibfnamefont{S.}~\bibnamefont{Inouye}},
  \bibinfo{author}{\bibfnamefont{T.}~\bibnamefont{Rosenband}},
  \bibnamefont{and} \bibinfo{author}{\bibfnamefont{W.}~\bibnamefont{Ketterle}},
  \bibinfo{journal}{Phys.~Rev.~Lett.} \textbf{\bibinfo{volume}{87}},
  \bibinfo{pages}{130402} (\bibinfo{year}{2001}).

\bibitem[{\citenamefont{Greiner et~al.}(2002)\citenamefont{Greiner, Mandel,
  Esslinger, H{\"a}nsch, and Bloch}}]{bloch02}
\bibinfo{author}{\bibfnamefont{M.}~\bibnamefont{Greiner}},
  \bibinfo{author}{\bibfnamefont{O.}~\bibnamefont{Mandel}},
  \bibinfo{author}{\bibfnamefont{T.}~\bibnamefont{Esslinger}},
  \bibinfo{author}{\bibfnamefont{T.}~\bibnamefont{H{\"a}nsch}},
  \bibnamefont{and} \bibinfo{author}{\bibfnamefont{I.}~\bibnamefont{Bloch}},
  \bibinfo{journal}{Nature} \textbf{\bibinfo{volume}{415}}, \bibinfo{pages}{39}
  (\bibinfo{year}{2002}).

\bibitem[{\citenamefont{St{\"o}ferle et~al.}(2004)\citenamefont{St{\"o}ferle,
  Moritz, Schori, K{\"o}hl, and Esslinger}}]{esslinger04}
\bibinfo{author}{\bibfnamefont{T.}~\bibnamefont{St{\"o}ferle}},
  \bibinfo{author}{\bibfnamefont{H.}~\bibnamefont{Moritz}},
  \bibinfo{author}{\bibfnamefont{C.}~\bibnamefont{Schori}},
  \bibinfo{author}{\bibfnamefont{M.}~\bibnamefont{K{\"o}hl}}, \bibnamefont{and}
  \bibinfo{author}{\bibfnamefont{T.}~\bibnamefont{Esslinger}},
  \bibinfo{journal}{Phys.~Rev.~Lett.} \textbf{\bibinfo{volume}{92}},
  \bibinfo{pages}{130403} (\bibinfo{year}{2004}).

\bibitem[{\citenamefont{Bloch}(2005)}]{bloch05}
\bibinfo{author}{\bibfnamefont{I.}~\bibnamefont{Bloch}},
  \bibinfo{journal}{Nature Physics} \textbf{\bibinfo{volume}{1}},
  \bibinfo{pages}{23} (\bibinfo{year}{2005}).

\bibitem[{\citenamefont{Hellweg et~al.}(2003)\citenamefont{Hellweg,
  Cacciapuoti, Kottke, Schulte, Sengstock, Ertmer, and Arlt}}]{ertmer03}
\bibinfo{author}{\bibfnamefont{D.}~\bibnamefont{Hellweg}},
  \bibinfo{author}{\bibfnamefont{L.}~\bibnamefont{Cacciapuoti}},
  \bibinfo{author}{\bibfnamefont{M.}~\bibnamefont{Kottke}},
  \bibinfo{author}{\bibfnamefont{T.}~\bibnamefont{Schulte}},
  \bibinfo{author}{\bibfnamefont{K.}~\bibnamefont{Sengstock}},
  \bibinfo{author}{\bibfnamefont{W.}~\bibnamefont{Ertmer}}, \bibnamefont{and}
  \bibinfo{author}{\bibfnamefont{J.~J.} \bibnamefont{Arlt}},
  \bibinfo{journal}{Phys.~Rev.~Lett.} \textbf{\bibinfo{volume}{91}},
  \bibinfo{pages}{010406} (\bibinfo{year}{2003}).

\bibitem[{\citenamefont{Walser}(2004)}]{walser04}
\bibinfo{author}{\bibfnamefont{R.}~\bibnamefont{Walser}},
  \bibinfo{journal}{Opt. Comm.} \textbf{\bibinfo{volume}{243}},
  \bibinfo{pages}{107} (\bibinfo{year}{2004}).

\bibitem[{\citenamefont{Kinoshita et~al.}(2005)\citenamefont{Kinoshita, Wenger,
  and Weiss}}]{Weiss05a}
\bibinfo{author}{\bibfnamefont{T.}~\bibnamefont{Kinoshita}},
  \bibinfo{author}{\bibfnamefont{T.}~\bibnamefont{Wenger}}, \bibnamefont{and}
  \bibinfo{author}{\bibfnamefont{D.~S.} \bibnamefont{Weiss}},
  \bibinfo{journal}{Phys.~Rev.~Lett.} \textbf{\bibinfo{volume}{95}},
  \bibinfo{pages}{190406} (\bibinfo{year}{2005}).

\bibitem[{\citenamefont{Ramsey}(1990)}]{ramsey90}
\bibinfo{author}{\bibfnamefont{N.}~\bibnamefont{Ramsey}},
  \bibinfo{journal}{Rev. Mod. Phys.} \textbf{\bibinfo{volume}{62}},
  \bibinfo{pages}{541} (\bibinfo{year}{1990}).

\bibitem[{\citenamefont{Kitagawa and Ueda}(1991)}]{ueda91}
\bibinfo{author}{\bibfnamefont{M.}~\bibnamefont{Kitagawa}} \bibnamefont{and}
  \bibinfo{author}{\bibfnamefont{M.}~\bibnamefont{Ueda}},
  \bibinfo{journal}{Phys.~Rev.~Lett.} \textbf{\bibinfo{volume}{67}},
  \bibinfo{pages}{1852} (\bibinfo{year}{1991}).

\bibitem[{\citenamefont{Holland and Burnett}(1993)}]{holland93}
\bibinfo{author}{\bibfnamefont{M.}~\bibnamefont{Holland}} \bibnamefont{and}
  \bibinfo{author}{\bibfnamefont{K.}~\bibnamefont{Burnett}},
  \bibinfo{journal}{Phys.~Rev.~Lett.} \textbf{\bibinfo{volume}{71}},
  \bibinfo{pages}{1355} (\bibinfo{year}{1993}).

\bibitem[{\citenamefont{Wineland et~al.}(1994)\citenamefont{Wineland,
  Bollinger, Itano, and Heinzen}}]{wineland94}
\bibinfo{author}{\bibfnamefont{D.~J.} \bibnamefont{Wineland}},
  \bibinfo{author}{\bibfnamefont{J.~J.} \bibnamefont{Bollinger}},
  \bibinfo{author}{\bibfnamefont{W.~M.} \bibnamefont{Itano}}, \bibnamefont{and}
  \bibinfo{author}{\bibfnamefont{D.~J.} \bibnamefont{Heinzen}},
  \bibinfo{journal}{Phys. Rev. A} \textbf{\bibinfo{volume}{50}},
  \bibinfo{pages}{67} (\bibinfo{year}{1994}).

\bibitem[{\citenamefont{Berman}(1997)}]{berman}
\bibinfo{author}{\bibfnamefont{P.}~\bibnamefont{Berman}},
  \emph{\bibinfo{title}{Atom interferometry}} (\bibinfo{publisher}{Academic
  Press}, \bibinfo{year}{1997}).

\bibitem[{\citenamefont{Vogel et~al.}(2006)\citenamefont{Vogel, Schmidt,
  Sengstock, Bongs, Lewoczko, Schuldt, Peters, Zoest, Ertmer, Rasel
  et~al.}}]{becmugrav06}
\bibinfo{author}{\bibfnamefont{A.}~\bibnamefont{Vogel}},
  \bibinfo{author}{\bibfnamefont{M.}~\bibnamefont{Schmidt}},
  \bibinfo{author}{\bibfnamefont{K.}~\bibnamefont{Sengstock}},
  \bibinfo{author}{\bibfnamefont{K.}~\bibnamefont{Bongs}},
  \bibinfo{author}{\bibfnamefont{W.}~\bibnamefont{Lewoczko}},
  \bibinfo{author}{\bibfnamefont{T.}~\bibnamefont{Schuldt}},
  \bibinfo{author}{\bibfnamefont{A.}~\bibnamefont{Peters}},
  \bibinfo{author}{\bibfnamefont{T.~V.} \bibnamefont{Zoest}},
  \bibinfo{author}{\bibfnamefont{W.}~\bibnamefont{Ertmer}},
  \bibinfo{author}{\bibfnamefont{E.}~\bibnamefont{Rasel}},
  \bibnamefont{et~al.}, \bibinfo{journal}{Appl. Phys B}
  \textbf{\bibinfo{volume}{84}}, \bibinfo{pages}{664} (\bibinfo{year}{2006}).

\bibitem[{\citenamefont{Nandi et~al.}(2006)\citenamefont{Nandi, Walser, Kajari,
  and Schleich}}]{nandi06}
\bibinfo{author}{\bibfnamefont{G.}~\bibnamefont{Nandi}},
  \bibinfo{author}{\bibfnamefont{R.}~\bibnamefont{Walser}},
  \bibinfo{author}{\bibfnamefont{E.}~\bibnamefont{Kajari}}, \bibnamefont{and}
  \bibinfo{author}{\bibfnamefont{W.~P.} \bibnamefont{Schleich}},
  \bibinfo{journal}{cond-mat/0610637}  (\bibinfo{year}{2006}).

\bibitem[{\citenamefont{Kajari et~al.}(2004)\citenamefont{Kajari, Walser,
  Schleich, and Delgado}}]{kajari04}
\bibinfo{author}{\bibfnamefont{E.}~\bibnamefont{Kajari}},
  \bibinfo{author}{\bibfnamefont{R.}~\bibnamefont{Walser}},
  \bibinfo{author}{\bibfnamefont{W.}~\bibnamefont{Schleich}}, \bibnamefont{and}
  \bibinfo{author}{\bibfnamefont{A.}~\bibnamefont{Delgado}},
  \bibinfo{journal}{Gen. Rel. Grav.} \textbf{\bibinfo{volume}{36}},
  \bibinfo{pages}{2289} (\bibinfo{year}{2004}).

\bibitem[{\citenamefont{Eckert et~al.}(2006)\citenamefont{Eckert, Hyllus,
  Bru\ss, Poulsen, Lewenstein, Jentsch, M\"uller, Rasel, and
  Ertmer}}]{ertmer06}
\bibinfo{author}{\bibfnamefont{K.}~\bibnamefont{Eckert}},
  \bibinfo{author}{\bibfnamefont{P.}~\bibnamefont{Hyllus}},
  \bibinfo{author}{\bibfnamefont{D.}~\bibnamefont{Bru\ss}},
  \bibinfo{author}{\bibfnamefont{U.~V.} \bibnamefont{Poulsen}},
  \bibinfo{author}{\bibfnamefont{M.}~\bibnamefont{Lewenstein}},
  \bibinfo{author}{\bibfnamefont{C.}~\bibnamefont{Jentsch}},
  \bibinfo{author}{\bibfnamefont{T.}~\bibnamefont{M\"uller}},
  \bibinfo{author}{\bibfnamefont{E.~M.} \bibnamefont{Rasel}}, \bibnamefont{and}
  \bibinfo{author}{\bibfnamefont{W.}~\bibnamefont{Ertmer}},
  \bibinfo{journal}{Phys. Rev. A} \textbf{\bibinfo{volume}{73}},
  \bibinfo{pages}{013814} (\bibinfo{year}{2006}).

\bibitem[{\citenamefont{Dimopoulos et~al.}(2007)\citenamefont{Dimopoulos,
  Graham, Hogan, and Kasevich}}]{kasevich07a}
\bibinfo{author}{\bibfnamefont{S.}~\bibnamefont{Dimopoulos}},
  \bibinfo{author}{\bibfnamefont{P.}~\bibnamefont{Graham}},
  \bibinfo{author}{\bibfnamefont{J.}~\bibnamefont{Hogan}}, \bibnamefont{and}
  \bibinfo{author}{\bibfnamefont{M.}~\bibnamefont{Kasevich}},
  \bibinfo{journal}{Phys.~Rev.~Lett.} \textbf{\bibinfo{volume}{98}},
  \bibinfo{pages}{111102} (\bibinfo{year}{2007}).

\bibitem[{\citenamefont{Li et~al.}(2007)\citenamefont{Li, Tuchman, Chien, and
  Kasevich}}]{kasevich07b}
\bibinfo{author}{\bibfnamefont{W.}~\bibnamefont{Li}},
  \bibinfo{author}{\bibfnamefont{A.}~\bibnamefont{Tuchman}},
  \bibinfo{author}{\bibfnamefont{H.-C.} \bibnamefont{Chien}}, \bibnamefont{and}
  \bibinfo{author}{\bibfnamefont{M.}~\bibnamefont{Kasevich}},
  \bibinfo{journal}{Phys.~Rev.~Lett.} \textbf{\bibinfo{volume}{98}},
  \bibinfo{pages}{040402} (\bibinfo{year}{2007}).

\bibitem[{\citenamefont{M\o{}lmer and S\o{}rensen}(1999)}]{moelmer99}
\bibinfo{author}{\bibfnamefont{K.}~\bibnamefont{M\o{}lmer}} \bibnamefont{and}
  \bibinfo{author}{\bibfnamefont{A.}~\bibnamefont{S\o{}rensen}},
  \bibinfo{journal}{Phys.~Rev.~Lett.} \textbf{\bibinfo{volume}{82}},
  \bibinfo{pages}{1835} (\bibinfo{year}{1999}).

\bibitem[{\citenamefont{Julsgaard et~al.}(2001)\citenamefont{Julsgaard,
  Kozhekin, and Polzik}}]{polzik01}
\bibinfo{author}{\bibfnamefont{B.}~\bibnamefont{Julsgaard}},
  \bibinfo{author}{\bibfnamefont{A.}~\bibnamefont{Kozhekin}}, \bibnamefont{and}
  \bibinfo{author}{\bibfnamefont{E.}~\bibnamefont{Polzik}},
  \bibinfo{journal}{Nature} \textbf{\bibinfo{volume}{413}},
  \bibinfo{pages}{400} (\bibinfo{year}{2001}).

\bibitem[{\citenamefont{S{\o}rensen et~al.}(2001)\citenamefont{S{\o}rensen,
  Duan, Cirac, and Zoller}}]{zoller_nature01}
\bibinfo{author}{\bibfnamefont{A.}~\bibnamefont{S{\o}rensen}},
  \bibinfo{author}{\bibfnamefont{L.-M.} \bibnamefont{Duan}},
  \bibinfo{author}{\bibfnamefont{J.~I.} \bibnamefont{Cirac}}, \bibnamefont{and}
  \bibinfo{author}{\bibfnamefont{P.}~\bibnamefont{Zoller}},
  \bibinfo{journal}{Nature} \textbf{\bibinfo{volume}{409}}, \bibinfo{pages}{63}
  (\bibinfo{year}{2001}).

\bibitem[{\citenamefont{Kitagawa and Ueda}(1993)}]{ueda93}
\bibinfo{author}{\bibfnamefont{M.}~\bibnamefont{Kitagawa}} \bibnamefont{and}
  \bibinfo{author}{\bibfnamefont{M.}~\bibnamefont{Ueda}},
  \bibinfo{journal}{Phys.~Rev.~A} \textbf{\bibinfo{volume}{47}},
  \bibinfo{pages}{5138} (\bibinfo{year}{1993}).

\bibitem[{\citenamefont{You}(2003)}]{you03}
\bibinfo{author}{\bibfnamefont{L.}~\bibnamefont{You}},
  \bibinfo{journal}{Phys.~Rev.~Lett.} \textbf{\bibinfo{volume}{90}},
  \bibinfo{pages}{30402} (\bibinfo{year}{2003}).

\bibitem[{\citenamefont{Javanainen and Ivanov}(1999)}]{ivanov1999}
\bibinfo{author}{\bibfnamefont{J.}~\bibnamefont{Javanainen}} \bibnamefont{and}
  \bibinfo{author}{\bibfnamefont{M.~Y.} \bibnamefont{Ivanov}},
  \bibinfo{journal}{Phys. Rev. A} \textbf{\bibinfo{volume}{60}},
  \bibinfo{pages}{2351} (\bibinfo{year}{1999}).

\bibitem[{\citenamefont{Leggett}(2001)}]{leggett401}
\bibinfo{author}{\bibfnamefont{A.}~\bibnamefont{Leggett}},
  \bibinfo{journal}{Rev. Mod. Phys.} \textbf{\bibinfo{volume}{73}},
  \bibinfo{pages}{307} (\bibinfo{year}{2001}).

\bibitem[{\citenamefont{Gati et~al.}(2006)\citenamefont{Gati, Esteve,
  Hemmerling, Ottenstein, Appmeier, Weller, and Oberthaler}}]{oberthaler06b}
\bibinfo{author}{\bibfnamefont{R.}~\bibnamefont{Gati}},
  \bibinfo{author}{\bibfnamefont{J.}~\bibnamefont{Esteve}},
  \bibinfo{author}{\bibfnamefont{B.}~\bibnamefont{Hemmerling}},
  \bibinfo{author}{\bibfnamefont{T.}~\bibnamefont{Ottenstein}},
  \bibinfo{author}{\bibfnamefont{J.}~\bibnamefont{Appmeier}},
  \bibinfo{author}{\bibfnamefont{A.}~\bibnamefont{Weller}}, \bibnamefont{and}
  \bibinfo{author}{\bibfnamefont{M.}~\bibnamefont{Oberthaler}},
  \bibinfo{journal}{N. J. Phys} \textbf{\bibinfo{volume}{8}},
  \bibinfo{pages}{189} (\bibinfo{year}{2006}).

\bibitem[{\citenamefont{Poulsen and M\o{}lmer}(2002)}]{molmer02}
\bibinfo{author}{\bibfnamefont{U.~V.} \bibnamefont{Poulsen}} \bibnamefont{and}
  \bibinfo{author}{\bibfnamefont{K.}~\bibnamefont{M\o{}lmer}},
  \bibinfo{journal}{Phys. Rev. A} \textbf{\bibinfo{volume}{65}},
  \bibinfo{pages}{033613} (\bibinfo{year}{2002}).

\bibitem[{\citenamefont{Peters et~al.}(1999)\citenamefont{Peters, Chung, and
  Chu}}]{peters99}
\bibinfo{author}{\bibfnamefont{A.}~\bibnamefont{Peters}},
  \bibinfo{author}{\bibfnamefont{K.}~\bibnamefont{Chung}}, \bibnamefont{and}
  \bibinfo{author}{\bibfnamefont{S.}~\bibnamefont{Chu}},
  \bibinfo{journal}{Nature} \textbf{\bibinfo{volume}{400}},
  \bibinfo{pages}{849} (\bibinfo{year}{1999}).

\bibitem[{\citenamefont{Barone and Paterno}(1982)}]{barone82}
\bibinfo{author}{\bibfnamefont{A.}~\bibnamefont{Barone}} \bibnamefont{and}
  \bibinfo{author}{\bibfnamefont{G.}~\bibnamefont{Paterno}},
  \emph{\bibinfo{title}{Physics and Application of the Josephson Effect}}
  (\bibinfo{publisher}{Wiley Interscience}, \bibinfo{address}{New York},
  \bibinfo{year}{1982}).

\bibitem[{\citenamefont{Leggett and Sols}(1991)}]{leggett91}
\bibinfo{author}{\bibfnamefont{A.}~\bibnamefont{Leggett}} \bibnamefont{and}
  \bibinfo{author}{\bibfnamefont{F.}~\bibnamefont{Sols}},
  \bibinfo{journal}{Found. Phys.} \textbf{\bibinfo{volume}{21}},
  \bibinfo{pages}{353} (\bibinfo{year}{1991}).

\bibitem[{\citenamefont{Goldobin et~al.}(2005)\citenamefont{Goldobin, Vogel,
  Crasser, Walser, Schleich, Koelle, and Kleiner}}]{goldobin05}
\bibinfo{author}{\bibfnamefont{E.}~\bibnamefont{Goldobin}},
  \bibinfo{author}{\bibfnamefont{K.}~\bibnamefont{Vogel}},
  \bibinfo{author}{\bibfnamefont{O.}~\bibnamefont{Crasser}},
  \bibinfo{author}{\bibfnamefont{R.}~\bibnamefont{Walser}},
  \bibinfo{author}{\bibfnamefont{W.~P.} \bibnamefont{Schleich}},
  \bibinfo{author}{\bibfnamefont{D.}~\bibnamefont{Koelle}}, \bibnamefont{and}
  \bibinfo{author}{\bibfnamefont{R.}~\bibnamefont{Kleiner}},
  \bibinfo{journal}{Phys. Rev. B} \textbf{\bibinfo{volume}{72}},
  \bibinfo{pages}{054527} (\bibinfo{year}{2005}).

\bibitem[{\citenamefont{Braun}(1993)}]{braun93}
\bibinfo{author}{\bibfnamefont{P.}~\bibnamefont{Braun}},
  \bibinfo{journal}{Rev.~Mod.~Phys.} \textbf{\bibinfo{volume}{65}},
  \bibinfo{pages}{115} (\bibinfo{year}{1993}).

\end{thebibliography}
\end{document}